\documentclass{kluwer}    

\usepackage[dvips]{graphicx}

\newdisplay{guess}{Conjecture}

\begin{document}                                                               

\begin{article}
\begin{opening}         
\title{Linear and Circular Polarization Properties of Jets
\thanks{Invited talk}} 
\author{Mateusz \surname{Ruszkowski}}  
\runningauthor{Mateusz Ruszkowski}
\runningtitle{Linear and Circular Polarization Properties of Jets}
\institute{JILA, University of Colorado at Boulder, CO 80309, USA}
\date{September 30, 2002}

\begin{abstract}
I discuss the transfer of polarized synchrotron radiation in relativistic 
jets. I argue that the main mechanism responsible for the circular polarization
properties of compact synchrotron sources is likely to be
Faraday conversion and that, contrary to common expectation,
a significant rate of Faraday rotation does not necessarily
imply strong depolarization.
The long-term persistence of the sign of circular polarization, observed in
some sources,
is most likely due to a small net magnetic flux generated in the central
engine, carried
along the jet axis and superimposed on a highly turbulent magnetic field.
I show that the mean levels of circular and linear polarizations depend on
the number of field reversals along the line of sight and that the gradient
in Faraday rotation across turbulent regions can lead to
``correlation depolarization''.
The model is potentially applicable to a wide range of synchrotron sources.
In particular, I demonstrate how the model can naturally explain the
excess of circular over linear polarization in the Galactic Center (Sgr
A$^{*}$) and the low-luminosity AGN M81$^{*}$.
\end{abstract}


\end{opening}           

\section{Introduction}
Polarization has proven to be an important tool in AGN research.
In principle, linear and particularly circular polarization
observations of synchrotron radiation may permit
measurements of various properties of relativistic 
jets such as: magnetic field strength
and topology, the net magnetic flux carried by jets (and hence
generated in the central engine), the energy spectrum of radiating
particles,
and the jet composition (i.e., whether jets are mainly composed of
$e^{+}-e^{-}$ pairs or electron-proton plasma).
The renewed interest in polarization of compact radio sources stems from
two recent developments. 
First, \citeauthor{bow99} (1999) 
detected circular polarization using the Very Large 
Array (VLA) in
the best supermassive black hole candidate, the Galactic Center (Sgr
A$^{*}$). This discovery was quickly confirmed by \citeauthor{sau99} (1999) 
using the Australia Telescope Compact Array (ATCA).
Circular polarization was also detected in  
the celebrated X-ray binary system SS 433 \cite{fen00} and the microquasar
 GRS 1915 +105 \cite{fen02}.
Moreover, the Very Long Baseline Array (VLBA) has
now detected circular polarization in as many as 20 AGN \cite{war98,hom99}.
Second, it is now possible to measure circular
polarization with an unprecedented accuracy of 0.01\% using 
the ATCA \cite{ray00}.
This dramatic improvement
in the observational status of polarization measurements
 has also brought new questions.
For example, there is observational evidence that the sign of
circular polarization is persistent over decades in some sources
\cite{kom84,hom99,hug02}, which indicates that it
is a fundamental property of jets and strongly suggests that a small
magnetic flux is frozen into jets. Another problem, 
is how to reconcile the high level of circular
polarization with the lower value of linear polarization in Sgr A$^{*}$
\cite{bow99} and M81$^{*}$ \cite{bru01, bow02}.
Moreover, there is not even a general consensus on the mechanism
responsible for the circular polarization properties of jets \cite{war98}.

\section{Observational trends}
Compact radio sources typically show a linear polarization (LP) of a
few percent of the total intensity \cite{jon85}. This is much less than the
theoretical maximum for synchrotron sources, which can approach $70\%$ in
homogeneous sources with unidirectional magnetic field. Therefore, magnetic
fields in radio sources are believed to be highly inhomogeneous, although the
nonvanishing linear polarization is in itself an indirect indication of
a certain degree of ordering of the field. From the theoretical
point of view, ordered jet magnetic field is expected when
shocks compress an initially random field \cite{lai80,lai81,hug89}
or when such initial
fields are sheared along the jet \cite{lai80,lai81,beg84}. \\
\indent
Circular polarization (CP) is a common feature of quasars and blazars
\cite{ray00,hom01}. It is usually characterized by an approximately flat
spectrum, and is generated near synchrotron self-absorbed jet cores
\cite{hom99}.
CP is detected in about 30\%-50\% of these objects.
Measured degrees of CP are generally lower than the levels of LP
 and usually range between 0.1\% and 0.5\% \cite{hom99,hom01}. 
Observations of proper motion of CP-producing
regions in the quasar 3C 273 \cite{hom99} suggest that circular polarization is
intrinsic to the source, as opposed to being due to foreground effects.

\section{Mechanisms for producing circular polarization}
The most obvious candidate for generating circular
polarization in compact radio sources is intrinsic emission \cite{leg68}.
However, intrinsic CP will be strongly
suppressed by the tangled magnetic field and
possibly by $e^{+}-e^{-}$ pairs, which do not contribute CP. 
Other mechanisms have also been proposed, among which the most popular ones
are scintillation \cite{marm00},
general relativistic effects in dispersive plasma \cite{bb02} 
 and Faraday conversion \cite{jon77a,jon88,war98,rb02,bf02}.
The scintillation mechanism, in which circular
polarization is stochastically produced by a birefringent screen located
between the jet and the observer, fails to explain the persistent sign of 
circular polarization (if required by observations in a given source) 
as the time-averaged CP signal is predicted to vanish.
The second mechanism can be important close to the central black hole.
The last mechanism --- Faraday conversion --- is a very
promising one and in the next subsection I discuss it in more detail.

\subsection{Faraday rotation and conversion}
The polarization of radiation changes as it propagates through any
medium in which modes are characterized by different plasma speeds.
In the case of plasma dominated by cold electrons the modes are nearly 
circularly polarized. The
left and right circular modes have different phase velocities
and therefore the linear polarization vector of the propagating
radiation rotates. 
This effect -- Faraday rotation -- is a specific example of a more general
phenomenon 
called birefringence. In a medium whose natural modes are linearly or
elliptically polarized, such as a plasma of relativistic particles,
birefringence leads to the partial cyclic conversion between linearly and
circularly polarized radiation as the phase relationships between the modes
along the ray change with position.

\subsubsection{Strong rotativity}
Strong departures from
mode circularity occur only when radiation propagates within a small angle
$\sim\nu_{L}/\nu$ of the direction perpendicular to the magnetic
field, where $\nu_{L}=eB/2\pi m_{e}c$ (quasi-transverse limit, QT).
Therefore radiative transfer
is often performed in the quasi-longitudinal (QL) approximation.
In a typical observational situation it is usually assumed that Faraday
rotation within the source cannot be too large, as this will lead to the
suppression of linear
polarization. However, this constraint does not prevent rotativity from
achieving large values locally as long as the mean rotativity, i.e., averaged over
all directions of magnetic field along the line of sight,
is indeed relatively small. Such a situation may happen in a turbulent plasma.
Technically, the strong rotativity regime
($\zeta_{v}^{*2}\gg\zeta_{q}^{*2}$, see below) is equivalent to the QL limit
\cite{jon77b} and in this proceedings I mostly focus on this approximation.

\begin{figure}[t!]
\centerline{
\begin{tabular}{cc}
\includegraphics[width=3.5in]{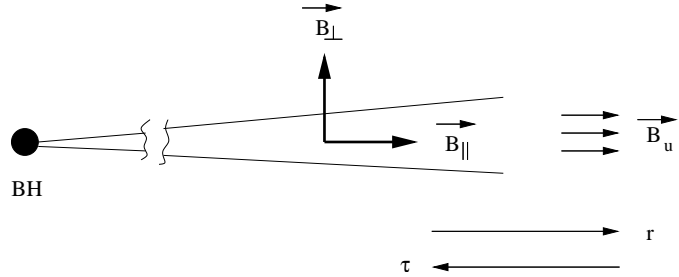}
\end{tabular}
\caption{Jet and its magnetic field (left panel) and geometry of the
assumed magnetic field assumed in the calculations (right panel).}
\label{}}
\end{figure}

\section{Model for polarization}
I consider a highly tangled magnetic field with a very small mean
component which is required to determine the sign of circular polarization.
From a theoretical view-point, we would expect some net poloidal magnetic
field, either originating from the central black hole or from the accretion
disk, to be aligned preferentially along the jet axis.
Specifically, from equipartition and flux freezing
arguments applied to a conical jet
\cite{bla79} we get $\langle B^{2}_{\|}\rangle^{1/2}\sim\langle
B^{2}_{\bot}\rangle^{1/2}\sim B_{\rm rms}\propto r^{-1}$ where
$r$ is the distance along the synchrotron emitting source and
the symbols $\|$
and $\bot$ refer to magnetic fields parallel and perpendicular to the jet axis,
respectively. From the flux-freezing
argument applied to the small parallel bias in the magnetic field we obtain
$\langle B_{\bot}\rangle\sim 0$ and $\langle B_{\|}\rangle\propto
r^{-2}\propto\delta B_{\rm rms}$, where $\delta\equiv 
B_{u}/B_{\rm rms}\ll 1$
is the ratio of the uniform and fluctuating components of the magnetic
field (see Figure 1).

\begin{figure}[t!]
\centerline{
\begin{tabular}{cc}
\includegraphics[width=2in, height=2in]{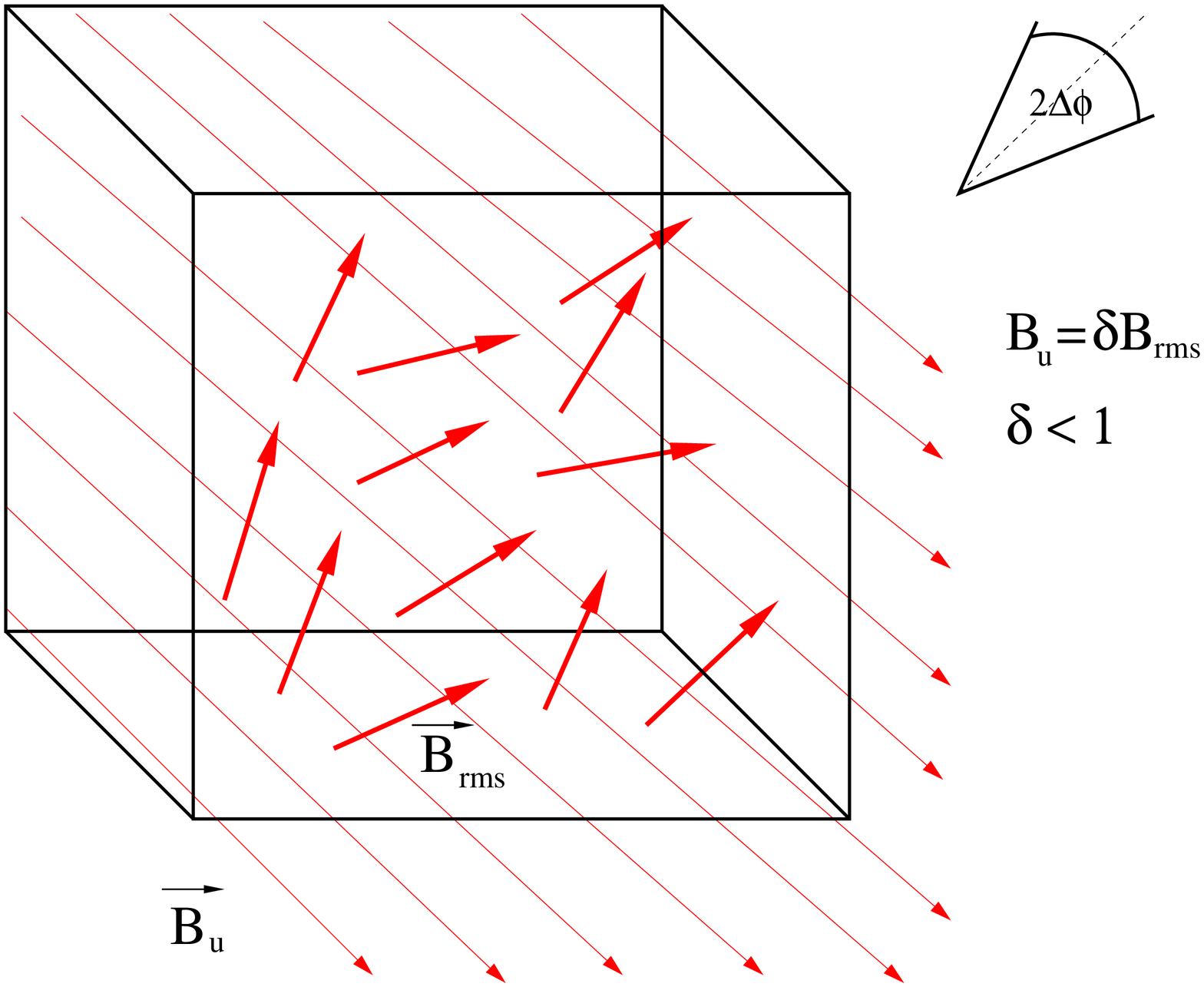}
\end{tabular}
\caption{Jet and its magnetic field (left panel) and geometry of the
assumed magnetic field assumed in the calculations (right panel).}
\label{}}
\end{figure}

\subsection{Mean Stokes parameters in the presence of field reversals}
I solve the radiative transfer of polarized radiation in a turbulent plasma
by adopting transfer equations for a piecewise homogeneous
medium with a weakly anisotropic dielectric tensor \cite{saz69,jon77a}. 
I assume that the mean rotation per unit synchrotron optical depth
$\langle\zeta_{v}^{*}\rangle\equiv\delta\zeta$ and that azimuthal angle 
$\phi\in [-\Delta\phi, \Delta\phi]$, 
polar angle $\theta\in[0,\pi]$, $\langle\sin2\phi\rangle =0$, and 
$\langle\cos2\phi\rangle =2p-1$, where $0\leq p\leq 1$ is a
parameter describing the polarization direction and degree of order in the
field (see Figure 2).
In the case when synchrotron depth is large and
the magnetic field unidirectional and appropriately rotated, 
the transfer equation has a particularly simple form:

\begin{equation}
\left(\begin{array}{cccc}
1 & \zeta_{q}  &  0 & 0 \\
\zeta_{q} & 1 &  \zeta_{v}^{*}    & 0 \\
0 & -\zeta_{v}^{*} & 1 &\zeta_{q}\\
0 & 0 & -\zeta_{q}^{*} & 1 
\end{array}\right)
\left(
\begin{array}{c}
I\\
Q\\
U\\
V
\end{array}
\right)
=
\left(
\begin{array}{c}
1\\
\epsilon_{q}\\
0\\
0
\end{array}
\right)
J,
\end{equation}

\noindent
where the orientation-dependent $J$ and 
$\zeta_{q}$, $\zeta_{q}^{*}$, $\zeta_{v}^{*}$ are the source function,
Q-absorptivity, convertibility and rotativity, respectively.
In a realistic situation, magnetic field will not be uniform. From the
analytic view-point,
polarization can be calculated by averaging equation of radiative transfer
over many orientations of magnetic field. Then, the levels of polarization
depend on the mean products of Stokes parameters and rotativity:

\begin{equation}
\langle\zeta_{v}^{*}U\rangle =\delta\zeta_{v}^{*}\langle U\rangle +
\langle \widetilde{\zeta}_{v}^{*}\widetilde{U}\rangle,
\end{equation}

\noindent
where the correlation term
$\langle\widetilde{\zeta}_{v}^{*}\widetilde{U}\rangle\propto N^{-1}$, 
where $N$ is
the number of field reversals along the line of sight. The correlation term
leads to depolarization when $N$ becomes sufficently small.
Heuristically, when $N$ is large then, as the radiation propagates through
turbulent zones, the polarization vector fluctuates to a lesser degree then
when $N$ is small. In the later case, projected orientation of the
polarization is more chaotic and the source is depolarized.

\subsection{Fiducial model}
I assume that the
typical mean Lorentz factor of radiating electrons $\gamma\sim 10^{2}$ and that
the electron energy distribution function has a power-law form
$n\propto\gamma^{-(2\alpha +1)}$, where $\alpha$ is the spectral index
of optically thin synchrotron emission. I use $\alpha=0.5$
and assume that the electron distribution is cut-off
below $\gamma_{i}\sim$ a few. For example, for the maximum brightness
temperature $T_{b}\sim 10^{11}K$ \cite{rea94} we have 
$\gamma\sim 3kT_{b}/m_{e}c^{2}\sim
50$, which corresponds to mean rotation and conversion per unit
synchrotron optical depth of order $\sim \delta\zeta_{v}^{*}\sim
3\times 10^{3}\delta\ln\gamma_{i}/\gamma_{i}^{3}$ and
$\zeta_{q}^{*}\sim -\ln(\gamma/\gamma_{i})$, respectively, for
$\nu\sim\gamma^{2}eB/2\pi m_{e}c$. Results from the simulations are shown
in Figure 4.

\begin{figure}[b!]
\centerline{
\begin{tabular}{cc}
\includegraphics[width=4in]{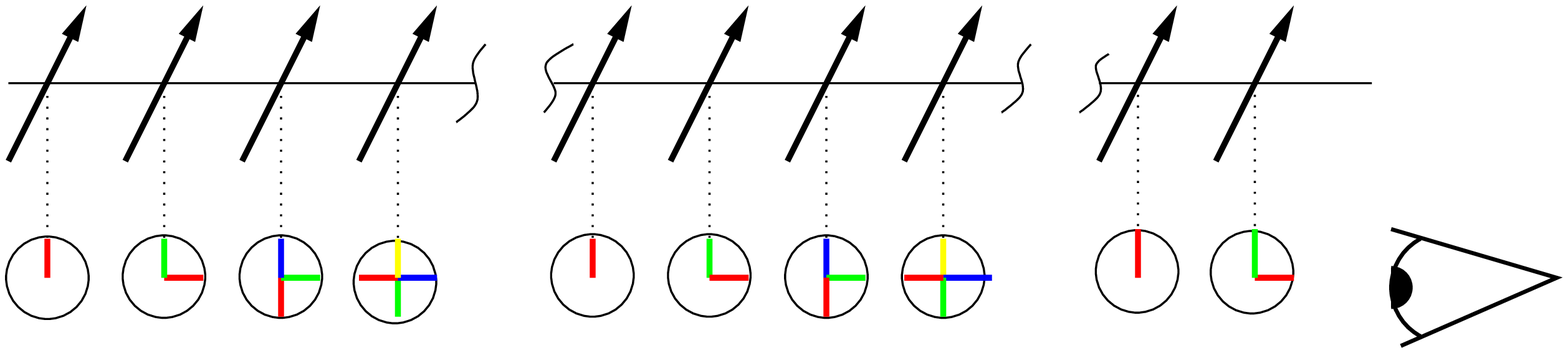}
\end{tabular}
\caption{This plot demonstrates how a uniform magnetic field can depolarize
most of the intrinsically polarized radiation when emission is present
along the line of sight.
Arrows indicate uniform magnetic field. Circular diagrams show the
orientation of the polarization vector. Locally generated linear
polarization is always denoted by a vertical line (i.e., 12 o'clock).
Linearly polarized flux from the neighboring regions cancels out. The
observed polarized radiation is generated in the regions closest to the
observer.}
\label{}}
\end{figure}

Note that, even though the rotation per unit depth is very large and the
magnetic field points (almost) randomly away and towards the observer,
the source is not totally depolarized. Note also that, 
circular polarization peaks around $\tau\sim 1$.
CP may change sign if $\tau$ varies strongly from optically thin to optically
thick regime. Linear polarization does tend to 
zero when the mean rotativity ($\propto\delta\propto B_{u}$) 
is high. Thus CP, which is produced by conversion of LP, also decreases for
high $\delta$.   (see Figure 3 and its caption for additional explanation).

\begin{figure}[t!]
\centerline{
\begin{tabular}{cc}
\includegraphics[width=4in]{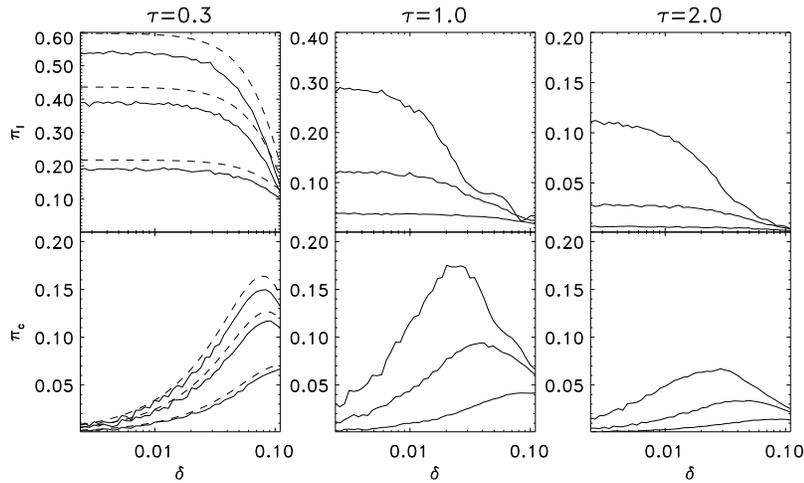}
\end{tabular}
\caption{Linear and circular polarization as a function of the ratio of the
uniform to fluctuating components of magnetic field $\delta
=B_{u}/B_{rms}$ for different synchrotron depths
(upper labels). Curves within each panel correspond to different numbers of
field reversals $N$ (for smaller $N$ polarization levels are lower).}
\label{cross_sections}}
\end{figure}

\subsection{Quasar 3C279}
\citeauthor{war98} (1998) 
reported the discovery of circular and linear polarization in
3C 279 and attributed CP to internal Faraday conversion.
Typical fractional linear and circular polarizations in 3C 279
are of order $\sim 10\%$ and $>1\%$, respectively.
They concluded that
if the jet is composed of normal plasma, then the low-energy cut-off of the
energy distribution of relativistic electrons must be as high as
$\gamma_{i}\sim 100$ in order to avoid Faraday depolarization and
overproduction of the jet kinetic power. 
They were unable to fit their polarization models to the observational data
for $\gamma_{i}>20$ and thus claimed that the jet must be
pair-dominated.
However, the above observational constraints on CP and LP and the
jet energetics can be satisfied for a variety of microscopic plasma
parameters. This is due to the fact that
different ``microscopic'' parameters, such as $\gamma_{i}$, the ratio of the cold to
relativistic electron number densities, or the positron fraction, can lead
to similar ``macroscopic'' parameters such as convertibility and rotativity.
In order to illustrate this, I consider two radically different examples
and show that both cases can lead to the same CP and LP.

\subsubsection{Electron-proton jet}
In this example, plasma is composed exclusively of a mixture of protons and electrons
with both relativistic (r) and cold (c) populations being present.
For a low-energy cut-off
$\gamma_{i}\sim 30$ and an electron number density-weighted
mean Lorentz factor $\gamma\sim 50$ and $\alpha =0.5$, we
get $\langle\zeta_{v}^{*}\rangle\sim 80(n_{c}/n_{r})\delta$ and
$\langle\zeta_{q}^{*}\rangle\sim -0.5$. For the above choice of parameters,
the main contribution to rotativity comes from
cold electrons as long as $n_{c}/n_{r}>5\times 10^{-3}$.
The required levels of LP and CP can be obtained, for example,
when $\tau=1$, $\alpha =0.5$, $N=15$, $2\Delta\phi =35^{o}$,
$\delta\sim 2.5\times 10^{-1}$ and $(n_{c}/n_{r})=7\times 10^{-2}$. 
Bear in mind that the admixture of cold electrons
does not have to be large to explain the data. 
Interestingly, a jet with such a plasma composition could carry
roughly as small a kinetic power as the pure
electron-positron jet with the same emissivity, since the ratio of kinetic powers of an
$e$-$p$ jet to a pure relativistic $e^{+}$-$e^{-}$ jet is
$\sim
18.4(\langle\gamma\rangle_{e^{+}e^{-}}/50)^{-1}(\gamma_{i,e^{+}e^{-}}/\gamma_{i,pe})$.

\subsubsection{Electron-positron jet}
The alternative possibility is that the jet is dominated
by relativistic pair plasma. For example, for $\gamma_{i}=2$ and
$\gamma =50$ we get $\langle\zeta_{q}^{*}\rangle\sim -3.1$ and
$\langle\zeta_{v}^{*}\rangle\sim 1.4\times 10^{2}\delta (n_{p}/n_{e-})$.
Agreement with the observed fractional LP and CP
can be obtained for example when the jet is pair-dominated
in the sense that $n_{e-}\gg n_{p}$, while being dominated dynamically by
protons. I was able to obtain the required polarization levels in this
case for $\tau =1$, $N=40$, $2\Delta\phi =30^{o}$, $\delta\sim
0.3$ and $n_{p}/n_{e-} =0.1$. Recent theoretical work of
\cite{sik00} suggests that jets may be pair-dominated numberwise
but still dynamically dominated by protons.

\subsection{Galactic Center -- Sgr A$^{*}$}
Recently \cite{bow99} reported the detection of circular polarization
from our Galactic Center (Sgr A$^{*}$) 
with the VLA, which was confirmed by \cite{sau99}
using ATCA. The typical level of CP in their observations was $\sim 0.3\%$,
greater than the level of linear polarization. This result
may seem surprising in light of the strong limits on the ratio of CP to
LP in AGN where CP/LP is usually much less than unity.
However, as explained above, an excess of CP over LP can be explained easily in the
framework of our model.
Archival VLA data indicate that the mean CP was stable over ten
years \cite{bow02}. This is also not surprising as our model naturally
predicts a persistent CP sign.
The average CP spectrum was characterized by a flat to slightly positive spectral index
($\pi_{c}\propto \nu^{\beta}, \beta >0$).
This result can also be accounted for in our model.
For example, in the framework of a self-absorbed,
self-similar jet model \cite{bla79} but with a small bias $\delta$
we have $\langle\zeta_{v}^{*}\rangle\propto\delta\propto B\propto\nu\propto
r^{-1}$ (see Figure 1).
The minimum and maximum size of Sgr A$^{*}$
constrain the brightness temperature to be
$10^{10}<T_{b}<5\times 10^{11}K$ \cite{mel01}, which is within the
range of typical AGN radio cores.
Taking $T_{b}\sim 10^{11}$ as the representative rest frame
value \cite{rea94}, we get $\gamma\sim 50$.
The required levels of CP and LP can be approximatelly obtained, for
example, for 
$\tau =1$, $N=45$, $p=1$, $\alpha =0.5$, $\gamma_{i}=3$, $\gamma\sim 50$ 
and $\delta =0.35$ (see Figure 5).\\
\indent
It has recently been suggested that observations of linear polarization
can be used to constrain the accretion rate in Sgr A$^{*}$ and other
low-luminosity AGN \cite{ago00,qua00}.
These authors base their argument on the assumption
that the Faraday rotation measure has to be sufficiently small in order not
to suppress strong linear polarization at higher frequencies
\cite{ait00,bo02}. This assumption
places limits on density and magnetic field strength and leads to
very low accretion rates $\sim 10^{-8}$ to $10^{-9}M_{\odot}$ yr$^{-1}$.
High values of linear polarization at higher frequencies 
may originate closer to the central black hole, where the bulk of the LP
emitting material may no longer be in the form of a self-absorbed jet but 
rather in the form of an accretion disk.
As noted by \citeauthor{ago00} (2000) and \citeauthor{qua00} (2000),
the above accretion rate is inconsistent with an advection-dominated 
model for Sgr A$^{*}$,
which assumes that the accretion rate is of order
the canonical Bondi rate $\sim 10^{-4}$ to $10^{-5}M_{\odot}$ yr$^{-1}$.
However, strong rotation measure does not in principle
limit densities and magnetic fields provided that the
field has a small magnetic flux associated with it,
which is required to define the sign of circular polarization
(note that the rotation angle is then reduced $\delta^{-1}$ times).

\subsection{Low-luminosity AGN -- M81$^{*}$} 
Circular polarization was also detected in the compact radio jet of
the nearby spiral galaxy M81 \cite{bru01}. Their estimated values of CP were
$0.27\pm 0.06\pm 0.07\%$ at 4.8 GHz and $0.54\pm 0.06\pm 0.07\%$ at 8.4
GHz, where
errors are separated into statistical and systematic terms (see also Bower
et al. 2002 for more results on LLAGN). This suggests
that the CP spectrum is flat or possibly inverted. They also detected no
linear polarization at a level of $0.1\%$, indicating that the source has a
high circular-to-linear polarization ratio. The spectral index indicates
that this source is synchrotron self-absorbed
and we can apply the same approach as for Sgr A$^{*}$. 

\begin{figure}[t]
\centerline{
\begin{tabular}{cc}
\includegraphics[width=1.9in, height=2in, angle=270]{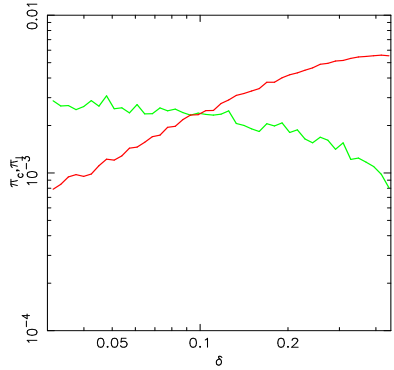}
\end{tabular}
\caption{Linear (decreasing curve) and circular polarization as a function
of $\delta$ for Sgr A$^{*}$ (intrinsic CP included)}
\label{}}
\end{figure}

\section{Conclusions}
I have considered the transfer of polarized synchrotron radiation in 
jets 
and have argued that Faraday conversion is the primary mechanism responsible
for the circular polarization properties of compact radio sources.
The modelis potentially applicable to a wide range of sources.
A crucial ingredient of the model is a small bias in the highly turbulent
magnetic field which accounts for the persistence of the sign of circular
polarization. This bias is a direct evidence for the net magnetic flux carried by
magnetically accelerated jets \cite{blap82,lic92}.\\
\indent
Extremely large Faraday rotation
per unit synchrotron absorption depth, does not necessarily lead to
depolarization provided that the mean rate of Faraday rotation across the
source is relatively small, or in other words, that the turbulent magnetic
field is characterized by a small magnetic flux. Indeed, a large Faraday
rotativity is required in order to explain the high ratio of circular to
linear polarization observed in some sources (Galactic Center, M81$^{*}$).
Constraints on jet composition or accretion rate, based
on the requirement that the source does not become Faraday depolarized,
may be circumvented under these conditions.\\
\indent
Gradients in Faraday rotation across turbulent
cells can lead to correlations between rotativity and Stokes $Q$ and $U$
parameters, which can result in ``correlation depolarization''.
Variations in the mean parameters are unlikely to change the helicity of
circular polarization unless a source undergoes a sharp transition from
very low to very high synchrotron depth.\\

\acknowledgements
I thank Mitch Begelman for collaboration on this project.
I also thank Don Backer, Roger Blandford, Geoff Bower, Avery Broderick, Rob Fender,
Dan Homan, Marek Sikora and John Wardle for insightful discussions.
I am also grateful to the organizers for organizing a very interesting conference.
This work was supported in part by NSF grant AST-9876887.

\bibliographystyle{klunamed.bst}
\bibliography{cpmr.bib}

\begin{thebibliography}{}

\bibitem[\protect\citeauthoryear{Agol}{2000}]{ago00} 
\newblock Agol, E. 2000, ApJ, 538, L121

\bibitem[\protect\citeauthoryear{Aitken et al.}{2000}]{ait00} 
\newblock Aitken, D.K., Greaves, J.,
Chrysostomou, A., Jennes, T., Holland, W., Hough, J.H., Pierce-Price, D,
and Richer, J. 2000, ApJ, 534, L173

\bibitem[\protect\citeauthoryear{Beckert and Falcke}{2002}]{bf02} 
\newblock Beckert, T., and Falcke, H. 2002, A\&A, 388, 1106

\bibitem[\protect\citeauthoryear{Begelman, Blandford and
Rees}{1984}]{beg84} 
\newblock Begelman, M.C., Blandford R.D., and Rees, M.J. 1984, Rev. Mod. Phys., 56, 255

\bibitem[\protect\citeauthoryear{Blandford and K\"{o}nigl}{1979}]{bla79} 
\newblock Blandford, R.D., and K\"{o}nigl, A. 1979, ApJ, 232, 34

\bibitem[\protect\citeauthoryear{Blandford and Payne}{1982}]{blap82} 
\newblock Blandford, R.D., and Payne, D.G. 1982, MNRAS, 199, 883

\bibitem[\protect\citeauthoryear{Bower et al.}{1999}]{bow99} 
\newblock Bower, G.C., Falcke, H., and Backer, D.C. 1999, ApJ, 523, L29

\bibitem[\protect\citeauthoryear{Bower et al.}{2000}]{bow00b} 
\newblock Bower, G.C., Falcke, H., Sault, H., and Backer, D.C. 2002, ApJ, 571, 843

\bibitem[\protect\citeauthoryear{Bower, Falcke and Mellon}{2002}]{bow02} 
\newblock Bower, G.C., Falcke, H., and Mellon, R.R. 2002, ApJL, in press

\bibitem[\protect\citeauthoryear{Bower et al.}{2002}]{bo02} 
\newblock Bower, G.C., Wright M.C.H., Falcke, H., and Backer, D.C. 2002, ApJ, submitted

\bibitem[\protect\citeauthoryear{Broderick and Blandford}{2002}]{bb02} 
\newblock Broderick, A., and Blandford, R., these proceedings

\bibitem[\protect\citeauthoryear{Brunthaler et al.}{2001}]{bru01}
\newblock Brunthaler, A., Bower, G.C., Falcke, H., and Mellon, R.R. 2001, ApJL, 560, 123

\bibitem[\protect\citeauthoryear{Fender et al.}{2000}]{fen00} 
\newblock Fender, R., Rayner D., Norris R., Sault R.J., 
and Pooley, G. 2000, ApJ, 530, L29

\bibitem[\protect\citeauthoryear{Fender et al.}{2002}]{fen02} 
\newblock Fender, R., et al. 2002, MNRAS, 336, 39

\bibitem[\protect\citeauthoryear{Homan and Wardle}{1999}]{hom99} 
\newblock Homan, D.C., and Wardle, J.F.C. 1999, AJ, 118, 1942

\bibitem[\protect\citeauthoryear{Homan et al.}{2001}]{hom01} 
\newblock Homan, D.C., Attridge, J.M., and Wardle J.F.C. 2001, ApJ, 556, 113

\bibitem[\protect\citeauthoryear{Hughes et al.}{1989}]{hug89} 
Hughes, P.A., Aller H.D., and Aller M.F. 1989, ApJ, 341, 54

\bibitem[\protect\citeauthoryear{Hughes}{2002}]{hug02} 
\newblock Hughes, P.A. 2002, these proceedidings

\bibitem[\protect\citeauthoryear{Jones}{1988}]{jon88} 
\newblock Jones, T.W. 1988, ApJ, 332, 678

\bibitem[\protect\citeauthoryear{Jones and O'Dell}{1977a}]{jon77a} 
\newblock Jones, T.W., and O'Dell S.L. 1977, ApJ, 214, 522

\bibitem[\protect\citeauthoryear{Jones and O'Dell}{1977b}]{jon77b} 
\newblock Jones, T.W., and O'Dell S.L. 1977, ApJ, 215, 236

\bibitem[\protect\citeauthoryear{Jones et al.}{1985}]{jon85} 
\newblock Jones, T.W., Rudnick, L., Aller, H.D.,
Aller, M.F., Hodge, P.E., and Fiedler, R.L. 1985, ApJ, 290, 627

\bibitem[\protect\citeauthoryear{Komesaroff et al.}{1984}]{kom84} 
\newblock Komesaroff, M.M., Roberts, J.A.,
Milne, D.K., Rayner, P.T., and Cooke, D.J. 1984, MNRAS, 208, 409

\bibitem[\protect\citeauthoryear{Laing}{1980}]{lai80} 
\newblock Laing, R.A. 1980, MNRAS, 193, 493

\bibitem[\protect\citeauthoryear{Laing}{1981}]{lai81} 
\newblock Laing, R.A. 1981, ApJ, 248, 87

\bibitem[\protect\citeauthoryear{Legg and Westfold}{1968}]{leg68}
\newblock Legg, M.P.C., and Westfold, K.C. 1968, ApJ, 154, 499

\bibitem[\protect\citeauthoryear{Li et al.}{1992}]{lic92} 
\newblock Li Z-Y., Chiueh, T., and Begelman M.C. 1992, ApJ, 394, 459

\bibitem[\protect\citeauthoryear{Macquart and Melrose}{2000}]{marm00} 
\newblock Macquart, J.-P., and Melrose, D.B. 2000, ApJ, 545, 798

\bibitem[\protect\citeauthoryear{Melia and Falcke}{2001}]{mel01} 
\newblock Melia, F., and Falcke, H. 2001, ARA\&A, 39, 309

\bibitem[\protect\citeauthoryear{Quataert and Gruzinov}{2000}]{qua00}
\newblock Quataert, E., and Gruzinov, A. 2000, ApJ, 545, 842

\bibitem[\protect\citeauthoryear{Rayner et al.}{2000}]{ray00} 
\newblock Rayner, D.P., Norris, R.P., and Sault, R.J. 2000, MNRAS, 319, 484

\bibitem[\protect\citeauthoryear{Readhead}{1994}]{rea94} 
\newblock Readhead, A.C.S. 1994, ApJ, 426, 51

\bibitem[\protect\citeauthoryear{Ruszkowski and Begelman}{2002}]{rb02} 
\newblock Ruszkowski, M., and Begelman M.C. 2002, ApJ, 573, 485

\bibitem[\protect\citeauthoryear{Sault and Macquart}{1999}]{sau99} 
\newblock Sault, R.J., and Macquart, J.-P. 1999, ApJ, 526, L85

\bibitem[\protect\citeauthoryear{Sazonov}{1969}]{saz69} 
\newblock Sazonov, V.N. 1969, Sov. Phys. JETP, 29, 578

\bibitem[\protect\citeauthoryear{Sikora and Madejski}{2000}]{sik00}
\newblock Sikora, M., and Madejski, G. 2000, ApJ, 534, 109

\bibitem[\protect\citeauthoryear{Wardle et al.}{1998}]{war98} 
\newblock Wardle, J.F.C., Homan, D.C., Ojha, R.,
and Roberts, D.H. 1998, Nature, 395, 457

\end{thebibliography}
\end{article}
\end{document}